\documentclass[preprint]{elsarticle}

\usepackage{hyperref}
\usepackage{amsfonts,amsmath,amssymb} 
\usepackage{graphicx,graphics,color}

\biboptions{sort&compress}
\journal{Physics of the Dark Universe}

\renewcommand{\(}{\begin{equation}}
\renewcommand{\)}{\end{equation}}
\newcommand{\bea}{\begin{eqnarray}}
\newcommand{\eea}{\end{eqnarray}}

\newcommand{\beq}{\begin{equation}}
\newcommand{\eeq}{\end{equation}}
\renewcommand{\(}{\begin{equation}}
\renewcommand{\)}{\end{equation}}

\usepackage{graphicx}
\usepackage{bm,natbib}
\usepackage{amsmath}
\usepackage[latin1]{inputenc}
\usepackage[english]{babel}
\usepackage[T1]{fontenc}
\usepackage{amssymb}
\usepackage{amsfonts}
\usepackage{epsfig}
\usepackage{colordvi}
\usepackage{psfrag}
\usepackage{color}
\usepackage{dcolumn}
\usepackage{multirow}
\usepackage{hyperref}
\usepackage{epstopdf}
\usepackage{bm}
\usepackage{times}
\hypersetup{
  colorlinks=true,        
  linkcolor=blue,         
  citecolor=magenta,      
}

\newcommand{\be}{\begin{equation}}
\newcommand{\ee}{\end{equation}}
\newcommand{\ba}{\begin{eqnarray}}
\newcommand{\ea}{\end{eqnarray}}
\newcommand{\bc}{\begin{center}}
\newcommand{\ec}{\end{center}}

\begin{document}
\bibliographystyle{article}
\begin{frontmatter}

\title{Noether symmetries in Interacting Quintessence Cosmology}

\vspace{0.4cm}


\date{\today}
\author[ad1,ad2]{Ester Piedipalumbo\corref{corauth}}
\cortext[corauth]{Corresponding author}
\ead{ester@na.infn.it}
\address[ad1]{Dipartimento di Fisica, Universit\`{a} degli Studi di Napoli Federico II, 
Compl. Univ. Monte S. Angelo, 80126 Naples, Italy}
\address[ad2]{I.N.F.N., Sez. di Napoli, Compl. Univ. Monte S. Angelo, Edificio 6, 
via Cinthia, 80126 - Napoli, Italy}

\author[ad1,ad2,ad3]{Mariafelicia~De Laurentis}
\ead{felicia@na.infn.it}
\address[ad3]{Lab. Theor. Cosmology,Tomsk State University of Control Systems and Radioelectronics(TUSUR), 634050 Tomsk, Russia}

\author[ad1,ad2,adn1,adn2]{Salvatore~Capozziello}
\ead{capozziello@na.infn.it}
\address[adn1]{Gran Sasso Science Institute, Viale F. Crispi, 7, I-67100, L'Aquila, Italy}
\address[adn2]{Tomsk State Pedagogical University, ul. Kievskaya, 60, 634061 Tomsk, Russia}

\begin{abstract}
The Noether Symmetry Approach is applied to interacting quintessence cosmology with the aim to search for exact solutions and  select scalar-field  self-interaction potentials.  It turns out that  the solutions found are compatible with the accelerated 
expansion of the Universe and  with  observational dataset, as the SNeIa
 Pantheon data. 
\end{abstract}

\end{frontmatter}
\section{Introduction}
\label{sec:Introduction}

Over the last decades, observational data revealed that the present Universe is experiencing an accelerated expansion 
driven by the so-called dark energy \cite{Riess07,Union2.1, PlanckXIII,scolnic2018}.
 Dark energy was first revealed observationally by examining the light from distant type Ia supernovae \citep{perl98, per+al99}. Actually, it turned out that a new form of energy (with negative pressure) is needed to accelerate the Hubble flow.  According to recent estimates, dark energy provides about 70\% of the total amount of matter-energy in the Universe, and only $5\%$ of the Universe is constituted by standard elements we ({\it i.e.} protons, neutrons, electrons, photons, neutrinos,  and even gravitational waves). The remaining $25\%$ is constituted by dark matter. The nature of dark energy and dark matter is yet unknown, and,  despite many proposals, both of them escape a final understanding at fundamental level.
 So far, some of the proposed models to account for dark energy include  non-vanishing cosmological constant,  potential energy of some scalar fields, 
effects connected to inhomogeneous distributions of matter and averaging procedures, and effects due to alternative/extended theories of gravity \cite{nesseris_fr, Cai, mareknmc1,nesseris, sergey,report,vasilis,ftester,cosmo_ester}.
In particular, among these last theories, there are the scalar-tensor theories of gravity that includes both  scalar fields and  tensor fields to represent gravitational  interaction \cite{valerio}. These theories arise in fundamental contexts like, for example,  
low-energy limits of Kaluza-Klein theories, string gravity or quantum field theory formulated in curved spacetimes \cite{Overduin,Birrell,report}. Furthermore,  they have been widely investigated in both early- and late-universe expansions \cite{Saha,Joyce2016, marekstar}. Recently, it  has been suggested that 
such  scalar fields are not fundamental but they might consist of fermion condensates \cite{VFC,CCFPV}. There are also other possibilities  listed and discussed, for example,  in  \cite{sante_interaction2018}.
Actually, the dynamical dark energy models have their own issues, as the cosmic coincidence problem, {\it i.e.}, the circumstance that  dark energy and dark matter are, today,  of the same order of magnitude
even if they evolve independently. Non-gravitational interactions between dark matter 
and dark energy has been recently proposed to avoid such an issue. Interacting dark matter/dark energy models have been studied 
in different contexts~\cite{dmde,amendola,cham,chamcosmo,wei,peebles04,neil,das,das15,Bonometto17, Bonometto19,marek1, mareknmc1,sanyal,sante_interaction2018, rocco2}.
However, the form of the interaction between dark matter and dark energy is not known, and there is an arbitrary  freedom to choose any specific interaction model.

This paper is aimed to understand whether this coupling can be connected to some symmetry.
Specifically,  by the so called {\it Noether Symmetry Approach}, it is possible to select suitable forms of interactions and scalar field potentials capable of addressing   the dark matter/dark energy issue in the cosmological expansion.

In Section \ref{Sec2}, we  
search for  the existence of Noether symmetries to the point-like  Lagrangian describing a
single scalar field  cosmological model coupled to  dark matter. 
We show that the existence of symmetry allows a coupled dark energy field and   provides an explicit 
form for the self-interaction potential leading the dark matter-dark energy interaction.
In Section \ref{Sec3}, we derive  general exact solutions  which naturally give rise to 
accelerated behaviors.
Section \ref{Sec4} is devoted  to compare the theoretical solutions with observations  using  the publicly available Pantheon SNeIa dataset. 
In Section \ref{conclusion} we draw conclusions. 
\section{ Noether symmetries for interacting quintessence}
\label{Sec2}
The Noether Symmetry Approach is a useful tool 
to find out exact solutions, in particular in cosmology, and select conserved quantities 
\cite{PHRVA,RNCIB,defelice, mareknmc2, leandros,felixGB,andronikos, phantom, sanyal_ester1,sanyal_ester2}. The existence of 
Noether symmetry allows to reduce the dynamical system that, in most cases,  results integrable. It is interesting to note that
the self-interacting scalar-field potentials,
\cite{PHRVA,RNCIB,leandros}, the couplings \cite{PHRVA,RNCIB},  or the overall structure of the
theory \cite{defelice}, can be related to the existence of the  symmetries ({\it i.e.}  the conserved
quantities). In this sense, the Noether
Symmetry Approach is also a criterion to select reliable
models (see \cite{defelice,kostas} for a discussion).

In the present case,
 let us  consider a 4-dimensional gravitational  action where a quintessential, minimally coupled scalar field interacts with  the dark matter component:
\begin{eqnarray}
&&{\cal A }=\int d^{4}x\sqrt{-g}\left[ \frac{1}{2} R+ \frac{1}{2} g^{\mu \nu} \phi_{;\mu ,\nu}-V(\phi)+\tilde{\mathcal{ L}}_m\left(g_{\mu \nu},\phi\right)\right]\,,
\label{action}
\end{eqnarray}
where $g$ is the determinant of the metric tensor $g_{\mu\nu}$ while  $V(\phi)$ is the self-interaction potential of the scalar field $\phi$. Furthermore,  we consider \begin{equation}
\tilde{\mathcal{ L}}_m\left(g_{\mu \nu},\phi\right)= \mathcal{ L }_m(g_{\mu \nu})+\mathcal{ L }^{int}_m(g_{\mu \nu},\phi),
\end{equation}  
where  $\mathcal{ L }_m(g_{\mu\nu})$ is the standard matter Lagrangian term and  ${\cal L }^{int}_m(g_{\mu\nu},\phi)$ is the interaction term. We are using physical units $(G_N = c =\hbar= 1)$.

\subsection{Cosmological equations }
Assuming  a spatially flat Friedmann-Robertson-Walker
cosmology, with metric  signature $(-,+,+,+)$,
the resulting point-like  Lagrangian, derived from Eq.\eqref{action}, can be written as
\begin{eqnarray}
{\cal L }= 3 a \dot{a}^2 - a^3  \left( \frac{\dot{\phi}^2}{ 2} - V(\phi)\right) + M a^{-3(\gamma -1)}+ M a^{-3(\gamma -1)}F(a,\phi)\,,\label{lagrangian}
\end{eqnarray}
where $a=a(t)$ is the scale factor of the Universe, and the constant $M$ is related to the today matter density $\rho_{m}^{0}$, 
where $1 \leq \gamma \leq 2$. 
The term $M a^{-3(\gamma -1)}F(a,\phi)$ describes the interaction between the dark matter-dark energy 
components. Here,  the dot represents the  derivative with respect to the cosmic time $t$.
In the following, we will limit our analysis to $\gamma=1$, corresponding to the dust case. 
Due to the presence of the interaction term in the Lagrangian, the  dark-matter energy density no longer evolves simply as $a^{-3}$ but instead scales as
\begin{equation}
\rho_{\rm DM} \sim \frac{1+F(a,\phi)}{a^3}\,.
\label{rhoDM}
\end{equation}
It is  possible to redefine $F(a,\phi)$ in order to absorb all terms, including the potential $V(\phi)$, into a single function:  
 \begin{equation}
 F(a,\phi) \rightarrow \tilde{F}(a,\phi) = a^3 V(\phi) + M a^{-3 (\gamma-1)}(1+F(a,\phi) )\,.
 \label{redefinition-REf}
 \end{equation}
  This choice has no effect in the search for Noether symmetries and the results are the same with only a single function being involved. However, the form used in the action (\ref{lagrangian}) offers the advantage to highlight  contributions from different terms (scalar field potential, matter etc.).
With these considerations in mind, the Friedmann equations are:
\begin{eqnarray}
&& 3 H^2=\frac {1}{2} \dot{\phi}^2+ V(\phi)+ \frac{M}{a^3}\left(1+F(a,\phi)\right)\label{energy}\,,\\
&&2\left(\frac{\ddot{a}}{a}\right)+ H^2 + \frac {1}{2} \dot{\phi}^2 - V(\phi)- \frac{M}{a^2}\frac{\partial F(a,\phi)}{\partial a}=0\label{addot}\,,\\
&&\ddot{\phi}+3H\dot{\phi}=-\frac{d V(\phi)}{d\phi}-\frac{M}{a^3}\frac{\partial F(a,\phi)}{\partial\phi}
\label{KGg}\,.
\end{eqnarray}
Eq.\eqref{KGg} for the scalar field is a generalized Klein-Gordon equation, 
and differs from the usual one by the last term, due to the interaction with the dark matter. 
If we want to find a {\it standard} form of the acceleration equation Eq.\eqref{addot} as:
\begin{equation}
6 \left(\frac{\ddot{a}}{a}\right)=-\left(\rho_{eff}+ 3 p_{eff}\right)\label{addot2}\,,
\end{equation}
then an effective pressure of the $\phi$-field is given by
\begin{equation}
p_{\phi}^{eff}= \frac {1}{2} \dot{\phi}^2- V(\phi)+ \frac{M }{3 a^2}\frac{\partial F(a,\phi)}{\partial a}\,. \label{fi-pressure}
\end{equation}
Analogously, it is possible to define an effective energy density of the scalar field by comparing Eq.\eqref{energy} with its {\it standard} form: 
\begin{equation}
3 H^2=\rho_m+\rho_{\phi}\,.
\end{equation}
It turns out that
\begin{equation}
\rho_{\phi}^{eff}= \rho_{\phi} +M a^{-3}F(a,\phi)\,, \label{fi-effdensity}
\end{equation}
where $\rho_{\phi}$ is the standard term:
\begin{equation}\label{fi-stdensity}
\rho_{\phi}=\frac {1}{2} \dot{\phi}^2+ V(\phi)\,.
\end{equation}
These two expressions  define an effective equation of state 
\begin{equation}
w_\phi^{eff}=\displaystyle \frac{p_{\phi}}{ \rho_{\phi}}=\frac{\frac {1}{2} \dot{\phi}^2
- V(\phi)+ \frac{M }{3 a^2}\frac{\partial F(a,\phi)}{\partial a}}{\frac {1}{2} \dot{\phi}^2+ V(\phi)+M a^{-3}F(a,\phi)}\,,
\end{equation}
which drives the time behavior of the model.
\subsection{The Noether Symmetry Approach}
In order to derive an analytical form for the self-interaction potential $V(\phi)$ and for the dark matter- dark energy coupling $F(a,\phi)$, we proceed as follows, relying upon the work in Refs. \cite{PHRVA,RNCIB}. Given the point 
Lagrangian in Eq.\eqref{lagrangian}, the configuration space is ${\cal Q}\equiv\{a,\phi\}$,
while the tangent bundle is $T{\cal Q}\equiv \{ a,\phi, \dot{ a},\dot{\phi}\}$. We look for
point symmetries which are transformations on the tangent space $T\cal Q$ derived from transformations on the
base space $\cal Q$. The infinitesimal generator of a point transformation reads as
\begin{equation}
X=\alpha\frac{\partial}{ \partial a}
+\beta\frac{\partial}{\partial \phi}
+\frac{d\alpha }{ dt}\frac{\partial}{\partial {\dot a}}
+\frac{d \beta}{ dt}\frac{\partial}{ \partial \dot \phi}\,,
\label{(2.2)}
\end{equation}
where
\begin{equation}
\frac{d \alpha}{ dt}={\dot \alpha}
=\frac{\partial \alpha}{\partial a}{\dot a}
+\frac{\partial \alpha}{\partial \phi}{\dot \phi}\,,
\label{(2.3)}
\end{equation}
\begin{equation}
\frac{d \beta}{dt}={\dot \beta}
=\frac{\partial \beta }{\partial a}{\dot a}
+\frac{\partial \beta }{\partial \phi}{\dot \phi}\,.
\label{(2.4)}
\end{equation}
Here $\alpha$ and $\beta$ are functions of $a(t),\phi(t)$.
The vector field $X$ on $T\cal Q$ is the complete lift \cite{RNCIB} of the vector field on $\cal Q$
\begin{equation}
Y=\alpha\frac{\partial}{ \partial a}
+\beta \frac{\partial }{\partial \phi}\,.
\label{(2.5)}
\end{equation}
The unknown functions $\alpha$ and $\beta$ are determined by requiring that the Lie derivative 
along $X$ of the point Lagrangian ${\cal L}$ is vanishing, {\it i.e.}
\begin{equation}
L_{X}{\cal L}=X {\cal L}=\alpha \frac{\partial {\cal L}}{\partial a}
+\beta \frac{\partial {\cal L}}{ \partial \phi}
+\frac{d \alpha }{ dt}\frac{\partial {\cal L}}{\partial {\dot a}}
+\frac{d \beta}{ dt}\frac{\partial {\cal L}}{ \partial {\dot \phi}}=0\,.
\label{(2.6)}
\end{equation}
This means that the Lagrangian is conserved  along the flow generated by the vector 
field $X$, {\it i.e.} Eq.\eqref{(2.6)} holds all over the tangent bundle of the configuration space.
According to \eqref{(2.2)}-\eqref{(2.4)}, Eq. \eqref{(2.6)} takes  the form
\begin{eqnarray}
&&3 {\dot a}^{2} \left(\alpha +2a \frac{\partial \alpha }{\partial a}\right) 
-{\dot \phi}^{2}\frac{a^{2}}{2}\left(3 \alpha+2a \frac{\partial \beta }{ \partial \phi}\right)
\nonumber \\ &&
+ {\dot a} \; {\dot \phi} \; a \left(6 \frac{\partial \alpha} {\partial \phi}
-a^{2}\frac{\partial \beta}{\partial a} \right) 
\nonumber \\
&&+ \left[3 \alpha a^{2} V(\phi) +\alpha M \frac{\partial F}{\partial a}
+\beta \left(a^{3}V'(\phi)+M \frac{\partial F } {\partial \phi}\right)\right]=0\,.
\label{(2.7)}
\end{eqnarray}
For this equation to be satisfied, the terms in round and square brackets must vanish
separately, which yields the system of first-order partial differential equations
\begin{equation}
\alpha+2a \frac{\partial \alpha }{\partial a}=0,
\label{(2.8)}
\end{equation}
\begin{equation}
3 \alpha +2a \frac{\partial \beta} { \partial \phi}=0,
\label{(2.9)}
\end{equation}
\begin{equation}
6 \frac{\partial \alpha}{\partial \phi}
-a^{2}\frac{\partial \beta} { \partial a}=0,
\label{(2.10)}
\end{equation}
\begin{equation}  
\left[3 \alpha a^{2} V(\phi) +\alpha M \frac{\partial F }{\partial a}
+\beta \left(a^{3}V'(\phi)+M \frac{\partial F } {\partial \phi}\right)\right]=0.
\label{(2.11)}
\end{equation}
Here the prime indicates the derivative with respect to the scalar field $\phi$. 
Eqs. \eqref{(2.8)}-\eqref{(2.10)} can be solved by the factorized ansatz
\begin{equation}
\alpha(a,\phi)=A_{1}(a)B_{1}(\phi), \; \;
\beta(a,\phi)=A_{2}(a)B_{2}(\phi).
\label{(2.12)}
\end{equation}
Eq. \eqref{(2.8)} becomes then a first-order ordinary differential equation for $A_{1}(a)$,
which is solved by 
\begin{equation}
A_{1}(a)=\frac{1}{\sqrt{a}},
\label{(2.13)}
\end{equation}
up to a multiplicative constant, and hence Eqs. \eqref{(2.9)} and \eqref{(2.10)} lead to
\begin{equation}
A_{2}(a)\frac{dB_{2}} {d\phi}=-\frac{3 } {2}a^{-\frac{3} {2}}B_{1}(\phi),
\label{(2.14)}
\end{equation}
\begin{equation}
\frac{d A_{2}} {d a}B_{2}(\phi)=6 a^{-\frac{5 } {2}}\frac{dB_{1}} {d\phi}.
\label{(2.15)}
\end{equation}
Therefore, we obtain 
\begin{equation}
A_{2}(a)=\kappa a^{-\frac{3}{2}},
\label{(2.16)}
\end{equation}
\begin{equation}
\frac{dB_{1}} {d\phi}=-\frac{\kappa}{ 4}B_{2}(\phi), \;
\frac{dB_{2}}{d\phi}=-\frac{3}{2\kappa}B_{1}(\phi),
\label{(2.17)}
\end{equation}
where $\kappa$ is a constant.
This leads  to a second-order equation for $B_{2}(\phi)$, from which we get
\begin{equation}
\alpha=\frac{A e^{\frac{1}{2} \sqrt{\frac{3}{2}} \phi }
+B e^{-\frac{1}{2} \sqrt{\frac{3}{2}}\phi }}{\sqrt{a}},
\label{(2.18)}
\end{equation}
\begin{equation}
\beta=\frac{\sqrt{6} e^{-\frac{1}{2} \sqrt{\frac{3}{2}} \phi } \left(B-A
e^{\sqrt{\frac{3}{2}} \phi }\right)}{a^{3/2}}\,,
\label{(2.19)}
\end{equation}
where the value $a=0$ has to be excluded.
 We can choose $A=0$ and  solve  Eq. \eqref{(2.11)} separating  the part involving the potential $V(\phi)$ and its first derivative. We find:
 \begin{equation}
V(\phi)= V_0 e^{-\sqrt{\frac{3}{2}} \phi }
\label{eqVphi}\,.
\end{equation}
 Therefore, the remaining symmetry condition implies that $F(a,\phi)=\mathcal{F}(\phi -6\ln{a})$, that is $\mathcal{F}$ is an arbitrary function of $(\phi -6\ln{a})$. If we select $\mathcal{F}=e^{-\frac{k \left(\phi -\sqrt{6} \log (a)\right)}{\sqrt{6}}}$ we find, from \eqref{(2.11)},
\begin{equation}
F(a,\phi)= Q a^k e^{-\frac{k \phi }{\sqrt{6}}}\,,\label{eqF}
\end{equation}
where $V_{0}$ and $Q$ are constants.
It is worth noticing that, in this way, the Noether symmetry selects the self-interaction potential and  coupling  which are often  assumed in literature for phenomenological reasons \cite{das,das15,boehmer1,boehmer2,Bonometto17,Bonometto19}. It turns out that the solution in Eq. \eqref{eqF} can be recovered if we factorize 
\begin{equation}
F(a,\phi)=F_1(a) F_2(\phi)\,.
\end{equation}
Moreover, for $k=0$, we recover a quintessence scalar field with the exponential self-interaction potential.
It is worth noticig that the other branch, $B=0$, leads again to infinite possibilities, being $F(a,\phi)=\mathcal{F}(\phi +6\ln{a})$ an arbitrary function. In the general case, when both $A,B \neq 0$, it is:
\begin{eqnarray}
&&V(\phi)=e^{-\sqrt{\frac{3}{2}} \phi } V_0 \left(B-A e^{\sqrt{\frac{3}{2}} \phi }\right)^2\,,
\label{Vgen}
\end{eqnarray} 
and 
   \begin{eqnarray}
&&F(a,\phi)=\mathcal{F}\left(\phi -\sqrt{\frac{2}{3}} \left(3 \log
   (a)+2 \log \left(B-A e^{\sqrt{\frac{3}{2}} \phi }\right)\right)\right)   
   \end{eqnarray}
   where the term in parentheses of $F(a,\phi)$ indicates  an arbitrary function of its argument.

   In the following, we limit our analysis to the solution given in Eqs. \eqref{eqVphi} and \eqref{eqF}\,
  which reproduce  potentials and coupling used in literature, for which we can find exact solutions of the cosmological equations. 
\section{Exact cosmological solutions}
\label{Sec3}
Once  the infinitesimal generator $X$ is found, 
it is  possible to find a change of variables $\{a,\phi\} \rightarrow\{u,v\}$,
such that one of them (say $v$, for example) is cyclic for the Lagrangian ${\cal L}$ in Eq. \eqref{lagrangian}, and the transformed Lagrangian produces, in general,  solvable equations. Integrating the system of equations
$i_{X} d u = 0$ and $i_X d v = 1$ (where $i_X d u$ and $i_X d v$
are the contractions between the vector field $X$ and the
differential forms $d u$ and $d v$, respectively), we obtain
\begin{eqnarray}
&& a =(u v )^{\frac{1}{3}}\,,\label{newvariable1}\\
&& \phi = -\sqrt{\frac{2}{3}} \log \left[\frac{u}{v}\right]\,.\label{newvariable2}
\end{eqnarray}
Under this transformation, the Lagrangian takes the form
\begin{equation}\label{newl}
{\cal L}= M+ M Q u^{\frac{2 k}{3}}+ V_0 u^2 + \frac{4\dot{v}\dot{u}  }{3 }\,,
\end{equation}
where  $v$ is cyclic. The conserved current gives
\begin{equation}
\Sigma=\frac{\partial \cal L}{\partial \dot{v}} = \frac{4}{3}\dot{u}\,,
\end{equation}
which can be trivially integrated to obtain 
\begin{equation}\label{ut}
u=\frac{3}{4}\Sigma t+ u_0.
\end{equation}
The energy condition $E_{\cal L}=0$ has been used to find $v$. It is easy to show that  the variable
$v$ satisfies the following differential equation:
\begin{equation}\label{eqv}
\dot{v}-\frac{V_0}{\Sigma}\left(u_0+\frac{3\Sigma}{4}t\right)^{2}
-\frac{M}{\Sigma}\left[1+Q\left(u_0+\frac{3\Sigma}{4}t\right)^{\frac{2 k}{3}} \right]=0\,.
\end{equation}

It is
\begin{eqnarray}
v=&&\frac{2^{-\frac{4}{3} (k+3)}}{9 (2 k+3) \Sigma ^2}\times\left[ 144 M Q \left(3 \Sigma  t+4 u_0\right)^{\frac{2 k}{3}+1}\right.\nonumber\\&&\left.+2^{4 k/3} (2 k+3) \left(144
   \Sigma  \left(M t+\Sigma  v_0\right)+V_0 \left(3 \Sigma  t+4 u_0\right){}^3\right)\right]\,.
\label{solkgen}
\end{eqnarray}

The substitution of the function $u$ in Eq. \eqref{newvariable1}, and the solution $v$ of the Eq.\eqref{newvariable2},
provides the explicit forms of scale factor and scalar field.
A family of solutions, parametrized by $k$, are found. 
In the next Section, we will discuss some special values of $k$. 
In order to obtain $v(t)$, we can solve  Eq.  \eqref{newvariable2} after substituting the value for $k$, or, alternatively,  start from  Eq. \eqref{solkgen}, and specify the value $k$. It follows, that the two solutions differ for a constant. In the next Subsection we will discuss some special cases. 
\subsection{The case $k=-3$}
Let us consider  the special case  $k=-3$.  Eqs.\eqref{ut} and \eqref{eqv}
read as
\begin{eqnarray}
u&=&\frac{3}{4}\Sigma t+ u_0\,, \label{uk6b}\\
v&=&-\frac{16 M Q}{9 \Sigma ^3 t+12 \Sigma ^2 u_0}+\frac{M t}{\Sigma }+\frac{V_0 \left(3
   \Sigma  t+4 u_0\right){}^3}{144 \Sigma ^2}+v_0\,.\nonumber\\
   \label{vk6}
\end{eqnarray} 
In order to obtain the scale factor $a(t)$, and the scalar field $\phi(t)$, we set 
$v_0=0$, and consider some constraints among the integration constants. For this purpose we 
impose the condition $a(0)=0$, and we set the present time at $t_0 = 1$. This fixes the 
time-scale according to the  age of the Universe. Because of our choice, the 
expansion rate $H(t)$ is dimensionless, so that our Hubble constant $\widehat{H}_0=H(t_0)$ 
is clearly of order $1$ and not (numerically) the same as the $H_0$ that is usually measured 
in ${\rm km s^{-1} Mpc^{-1}}$. Actually, $\widehat{H}_0$ fixes only the product $h\tau$, and 
depends on the integration constants. Therefore, it is possible to \textit{constrain} their 
range of variability, starting from $\widehat{H}_0 $.
We then set $a_0 = a(1) = 1$, and $\widehat{H}_0=H(1)$.  

\begin{figure}[!ht]
\includegraphics[width=\linewidth,clip]{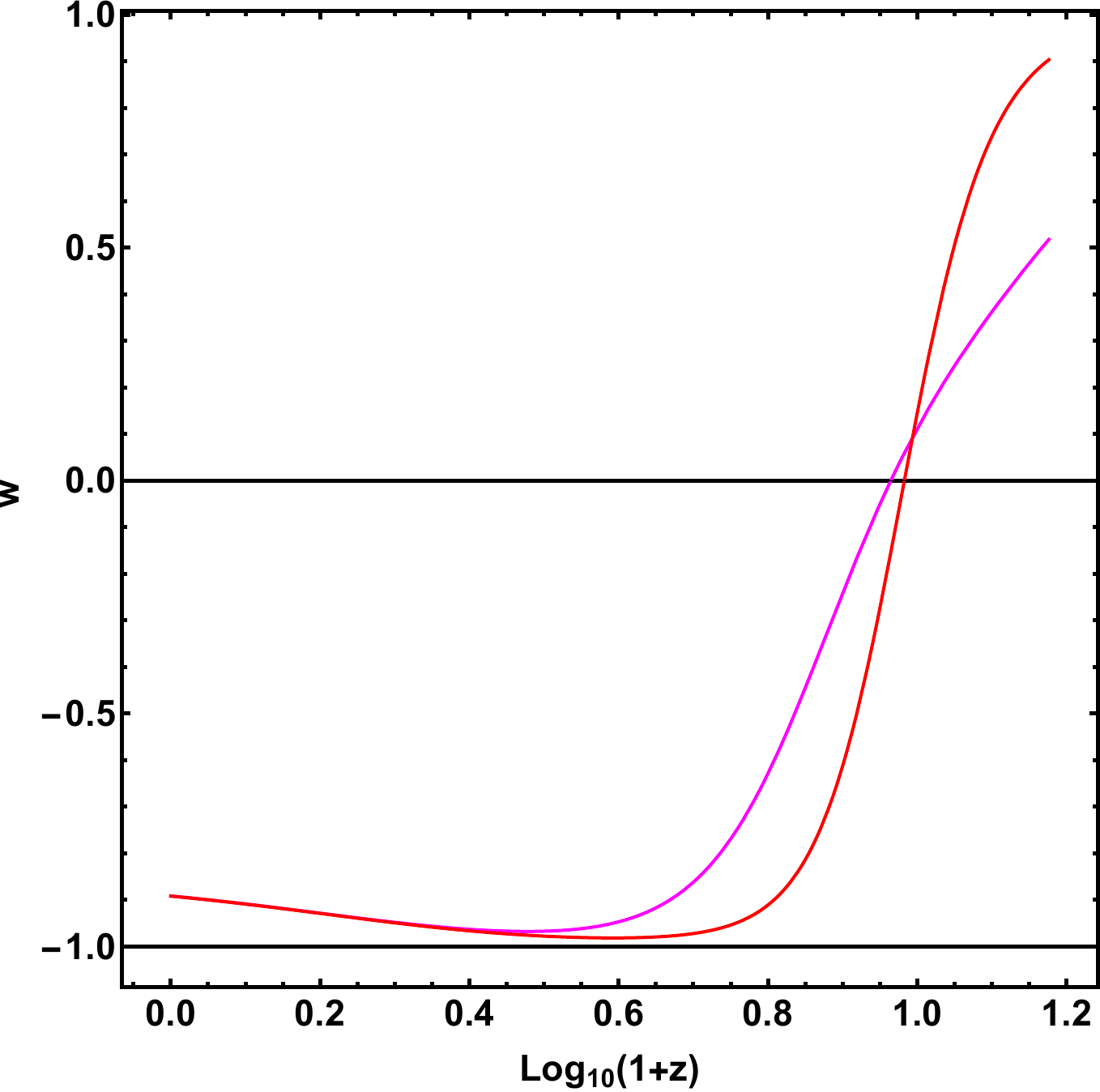}
\caption{The case k=-3: redshift dependence of the equation of state parameter $w_{eff}$ 
(red line) and  $w_\phi$ (magenta line) for fixed value of $H_0$, $u_0$, and $\Sigma$.}
\label{fig:wz}
\end{figure}

\begin{figure}
\centering
\includegraphics[width=\linewidth,clip]{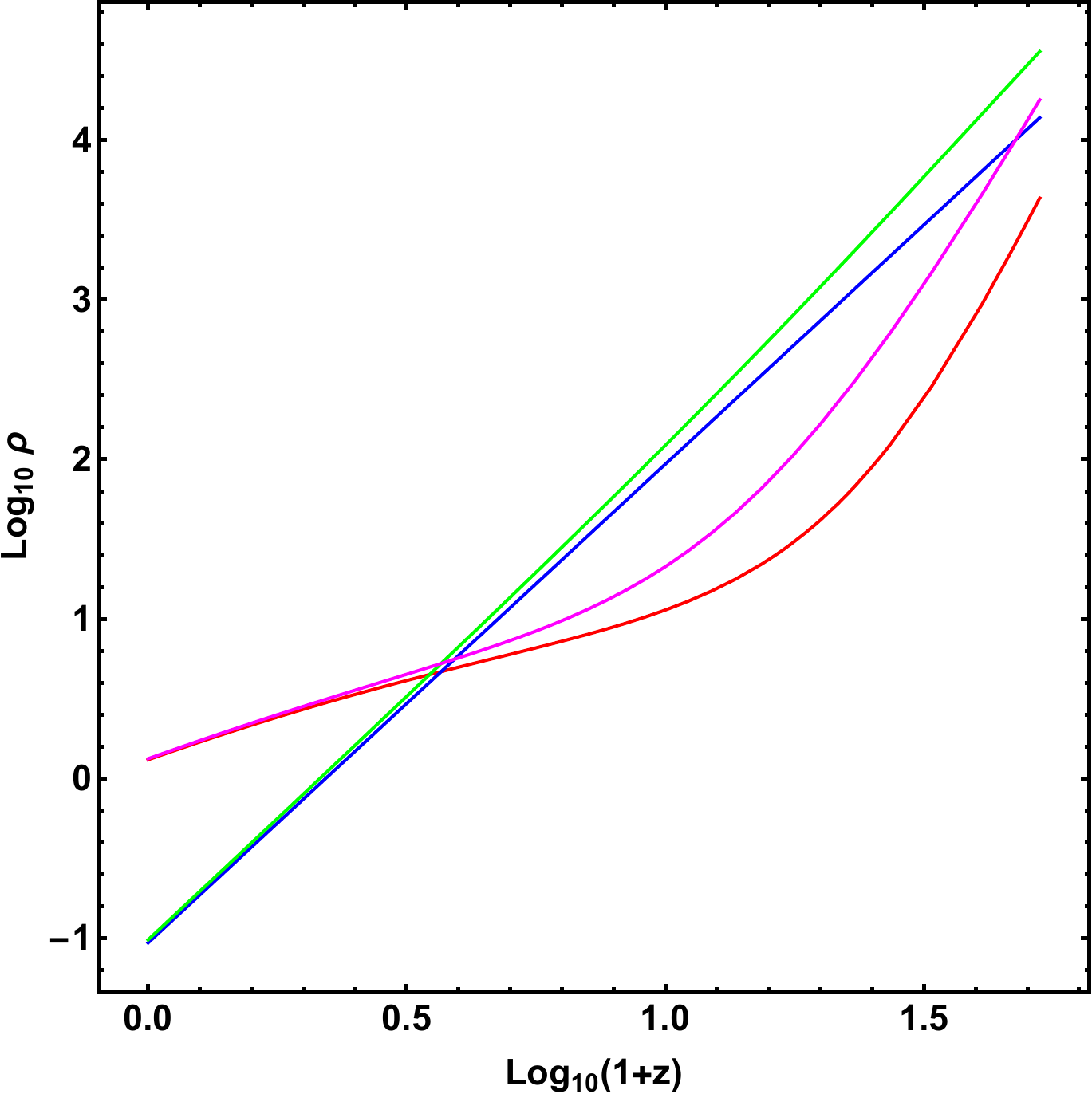}
\caption{The case $k=-3$: plot of $\log_{10}{\rho}$ versus ${\log}_{10} 1+z$ for fixed values of $H_0$, $u_0$ 
and $\Sigma$. The blue line indicates the log-log plot of $\log_{10}{\tilde{\rho}_{m}}\propto 
\log_{10}{a^{-3}} $, the green line is the dark matter  density $\rho_{\rm DM} \sim 
\displaystyle\frac{1+Q_1(a)Q_2(\phi)}{a^3}$, the red and magenta lines correspond to 
$\rho_{\phi}$ and $\rho_{eff}$ respectively, according to the Eqs.\eqref{fi-effdensity}
and \eqref{fi-stdensity}.}
\label{fig:logplot}
\end{figure}

By means of these choices and of the formulae  in Eqs. \eqref{eqaminus3} and \eqref{eqphiminus3}, we can construct all relevant cosmological parameters: $\rho_{\phi}$,$p_{\phi}$,$V_{\phi}$, $w_{\phi}$, 
$\rho^{eff}$, $p^{eff}$, and $w^{eff}$. 
\begin{figure}[!ht]
\includegraphics[width=\linewidth,clip]{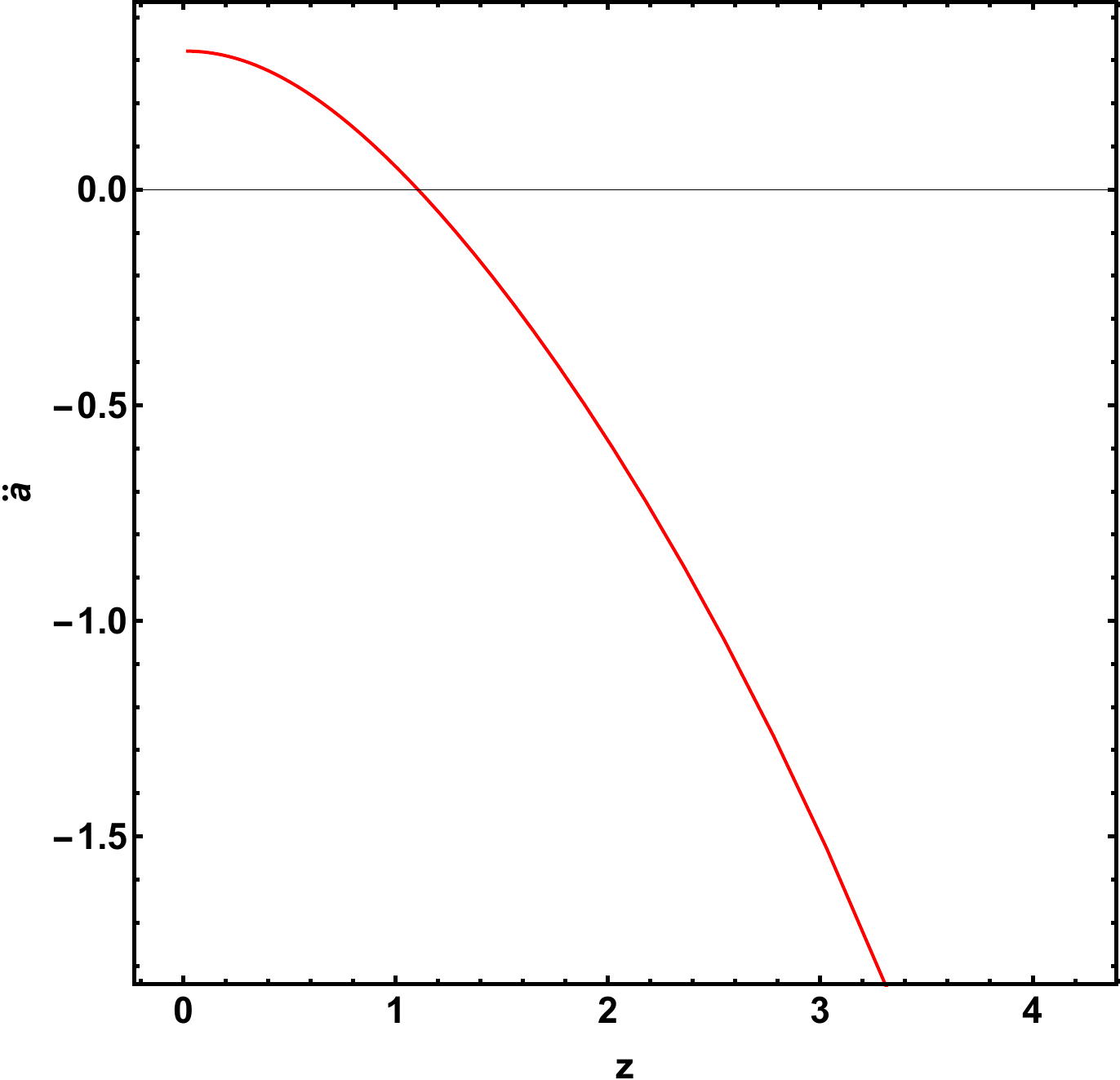}
\caption{The case $k=-3$: redshift dependence of the acceleration $\ddot{a}(t)$: we see that the 
model allows for an accelerated phase of expansion, as indicated by the observations.}
\label{fig:acc}
\end{figure}
As it is shown in Fig.(\ref{fig:acc}),  our model allows  an accelerated expansion as indicated from the observations and exhibits an evolving equation of state, as shown in Fig. (\ref{fig:wz}), where we compare the effective equation of state, $w_{eff}$, 
and the scalar field $w_{\phi}$. Furthermore, in Fig.(\ref{fig:logplot}),  we present the  plot $\log{\rho}$ - $\log (1+z)$ compared with  different matter-energy densities. Moreover, no violation of the weak energy condition is observed. 
\begin{figure}[!ht]
\includegraphics[width=\linewidth,clip]{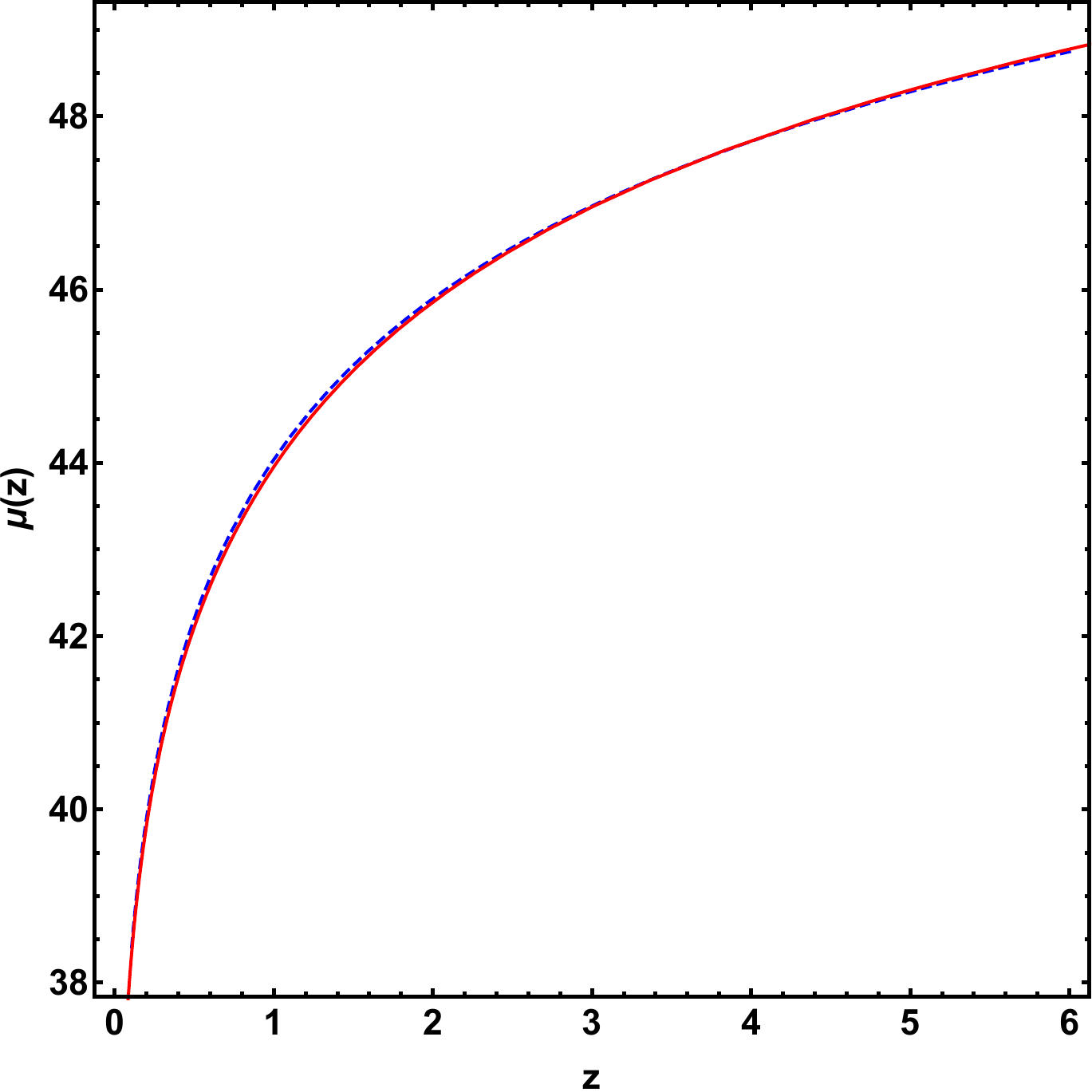}
\caption{The case $k=-3$:distance modulus for fixed values of $H_0$, $u_0$ (red solid line), compared with 
that one predicted in the standard $\Lambda$CDM model (blue dashed line).}
\label{modulus}
\end{figure}

In Fig. (\ref{modulus}),  we plot the distance modulus  for fixed values of $H_0$, 
$u_0$ and $\Sigma$, compared with that one predicted in the standard $\Lambda$CDM model.

\subsection{The cases $k=\frac{3}{2}$ and $k=3$.}
Another interesting solution is  for 
$k=\displaystyle\frac{3}{2}$. Also in this case we start from Eqs. (\ref{ut}) and (\ref{eqv}),  so that
\begin{eqnarray}
u&=&\frac{3}{4}\Sigma t+ u_0\,, \label{uk6}\\
v&=&\frac{24 M \left(\frac{1}{4} Q \left(3 \Sigma  t+4 u_0\right){}^2+6 \Sigma 
   t\right)+V_0 \left(3 \Sigma  t+4 u_0\right){}^3+144 \Sigma ^2 v_0}{144 \Sigma ^2}\nonumber \label{vk32}
\end{eqnarray} 

We then fix $v_0=0$,    $a_0 = a(1)= 1$, and $\widehat{H}_0=H(1)$ to get some constraints among the integration constants. The scale factor and the scalar fields are in Appendix.

\begin{figure}[!ht]
\includegraphics[width=\linewidth,clip]{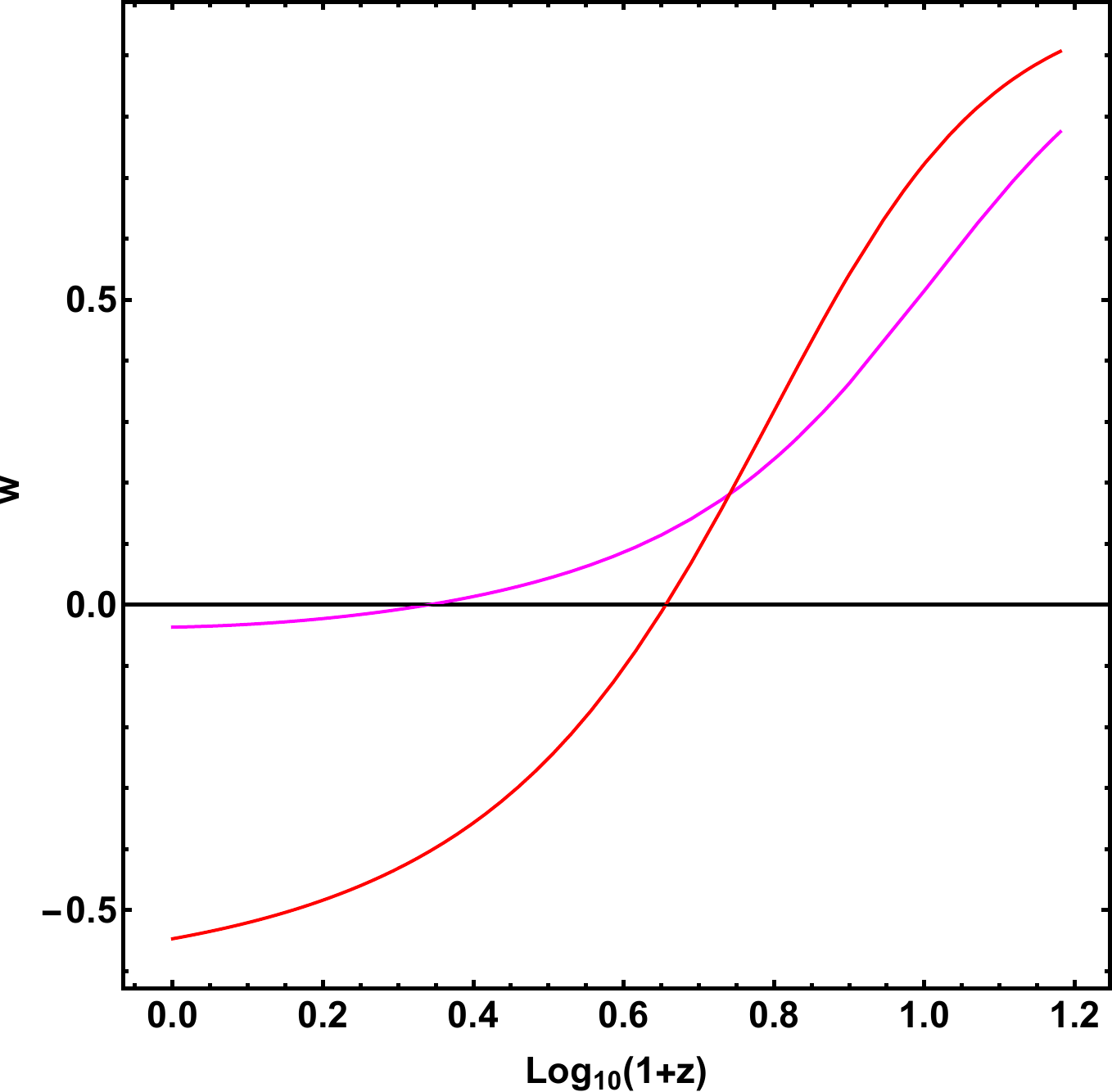}
\caption{The case $k=\frac{3}{2}$: redshift dependence of the equation of state parameter $w_{eff}$ (red line) and $w_\phi$ (magenta line) for fixed value of $H_0$, $u_0$, and $\Sigma$.}
\label{fig:wz3over2}
\end{figure}

\begin{figure}[!ht]
\includegraphics[width=\linewidth,clip]{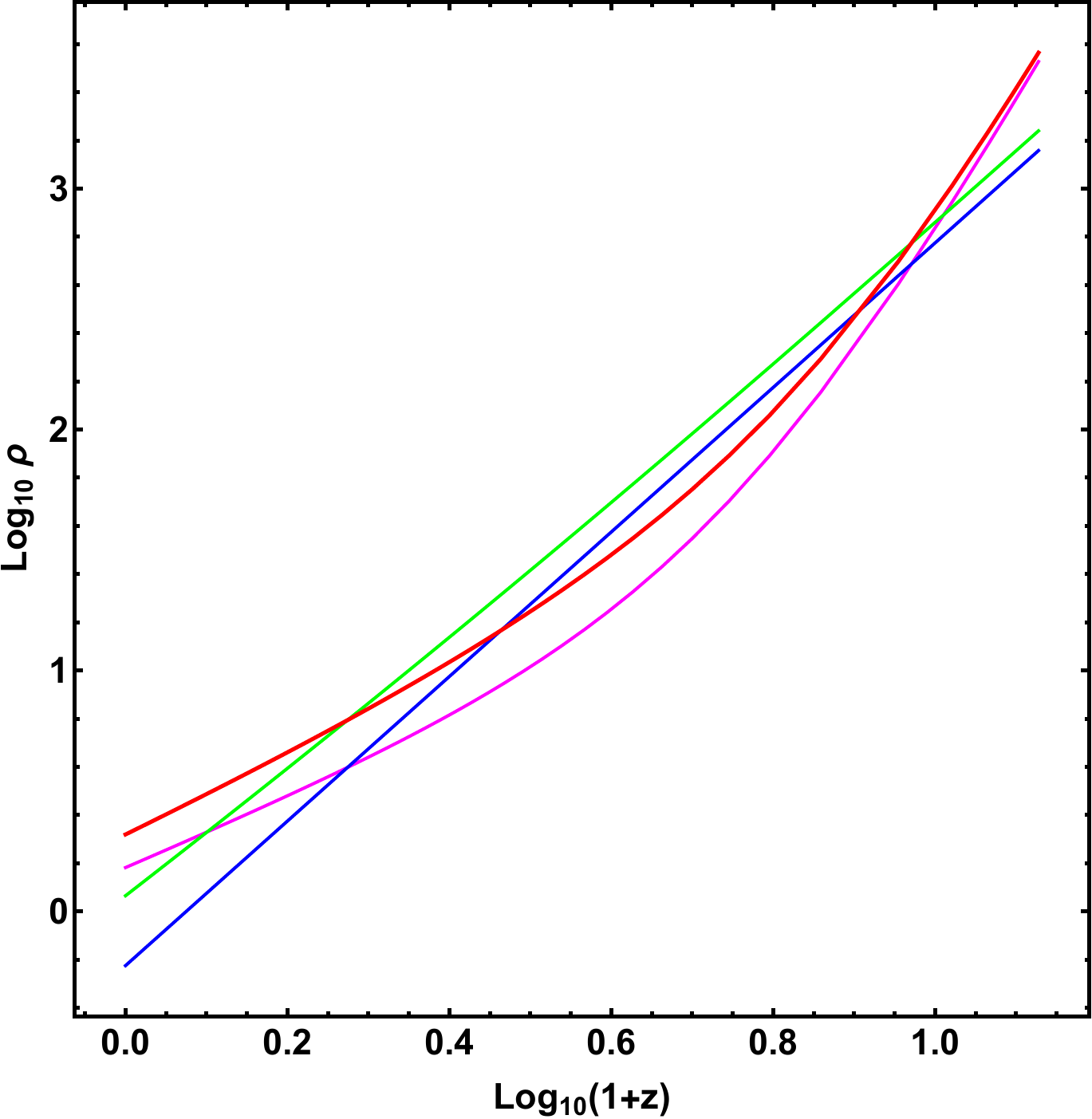}
\caption{$k=\frac{3}{2}$: plot of $\log_{10}{\rho}$ versus ${\log}_{10} (1+z)$ for fixed values of $H_0$, 
$u_0$ and $\Sigma$ in the $k=\displaystyle\frac{3}{2}$. As in the previous subsection the 
blue line indicates the log-log plot of $\log_{10}{\tilde{\rho}_{m}}\propto \log_{10}{a^{-3}} $, 
the green line is the dark matter energy density $\rho_{\rm DM} \sim \displaystyle
\frac{1+Q_1(a)Q_2(\phi)}{a^3}$, the red and magenta lines correspond to $\rho_{\phi}$ 
and $\rho_{eff}$ rispectively, according to the Eqs. \eqref{fi-effdensity} and \eqref{fi-stdensity}.}
\label{fig:logplot3over2}
\end{figure}

As it is shown in Fig. (\ref{fig:acc3over2}), also this solution allows for an accelerated expansion, 
and exhibits an evolving equation of state.  In Fig. (\ref{fig:wz3over2}),
 we compare the effective equation of state, $w_{eff}$, and that of  scalar field $w_{\phi}$. 
Finally, we present the  plot $\log{\rho}$ - $\log (1+z)$ 
compared with  different energy densities (\ref{fig:logplot3over2}). 
\begin{figure}[!ht]
 \includegraphics[width= \linewidth,clip]{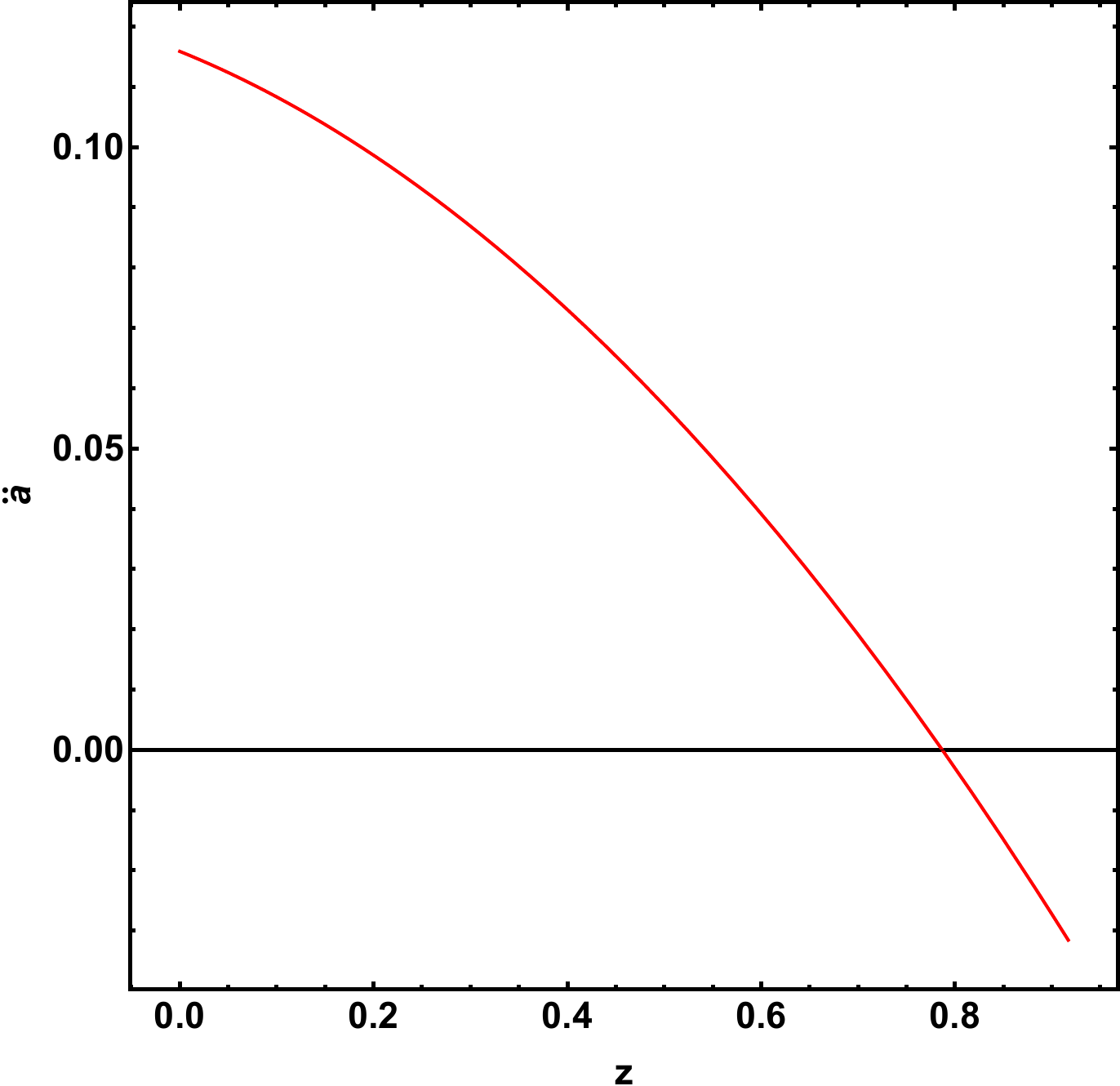}
\caption{The case $k=\frac{3}{2}$: redshift dependence of the acceleration $\ddot{a}(t)$: we see that also this model allows for an accelerated phase of expansion, as indicated by the observations.}
\label{fig:acc3over2}
\end{figure}
In particular, for $k=3$, setting $u_0=0$, $Q=\Sigma=1$, we recover the special case of a minimally coupled scalar field with the exponential potential already analyzed in \cite{marek1,phantom,esterexp}.
The scale factor, and the scalar field are given in \eqref{eqak3} and \eqref{eqphik3} (see Appendix).
\begin{figure}[!ht]
\includegraphics[width=\linewidth,clip]{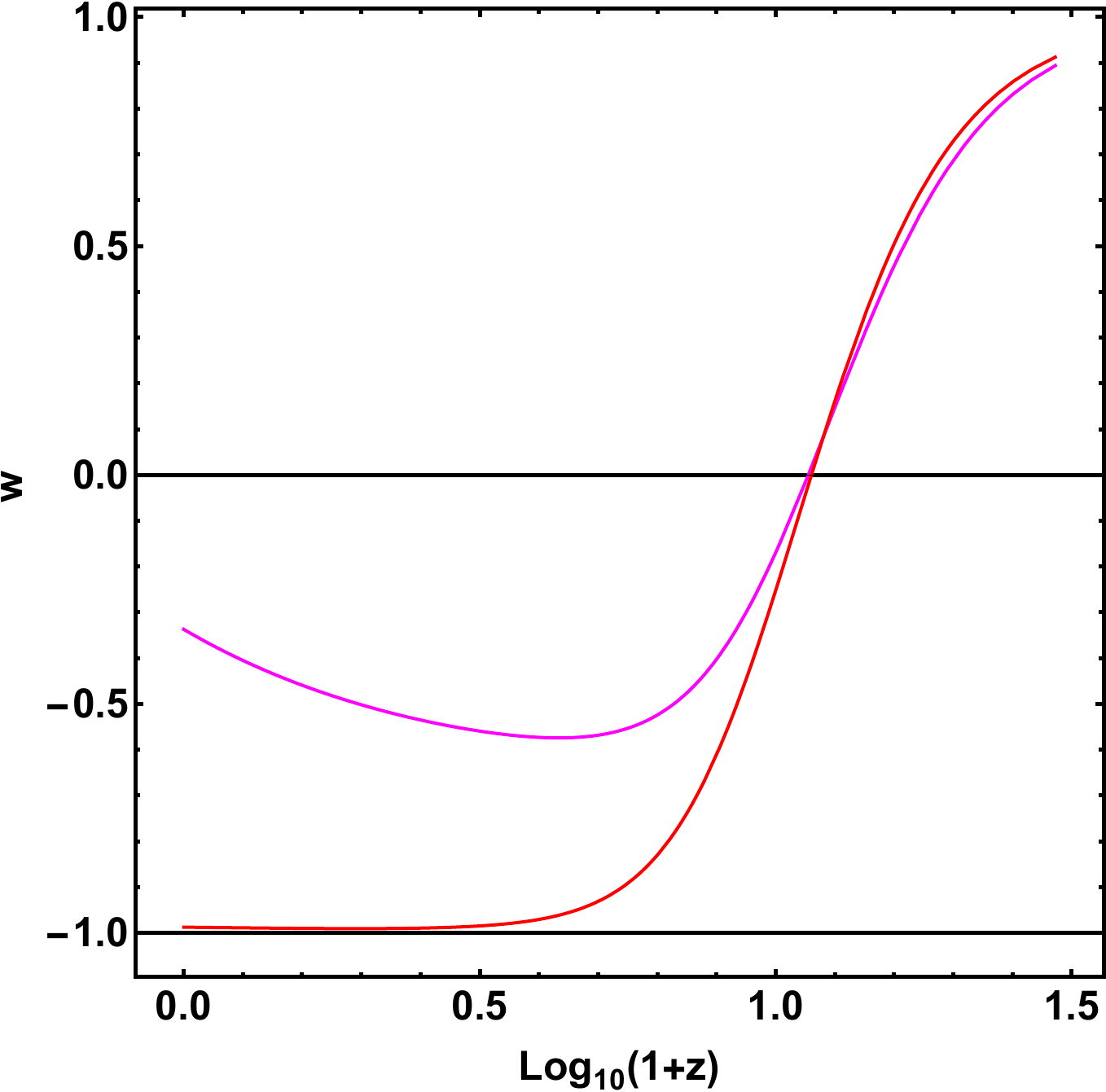}
\caption{The case $k=3$: redshift dependence of the equation of state parameter $w_{eff}$ (red line) and $w_\phi$ (magenta line) for fixed value of $H_0$, and $\Sigma$.}
\label{fig:wk3}
\end{figure}
Also this solution corresponds to an accelerated expansion of the Universe, and exhibits an evolving equation of state, as shown in Fig. (\ref{fig:wk3}).
\section{Comparing with  observations }
\label{Sec4}
In order to investigate the reliability of this class of models, we present a comparison of theoretical predictions with the most updated compilation of SNeIa given in the Pantheon survey \cite{scolnic2018}. Our  preliminary analysis is limited to the $k=3$ model.
\subsection{Data sample and statistical analysis}

\begin{figure}[!ht]
\includegraphics[width= \linewidth,clip]{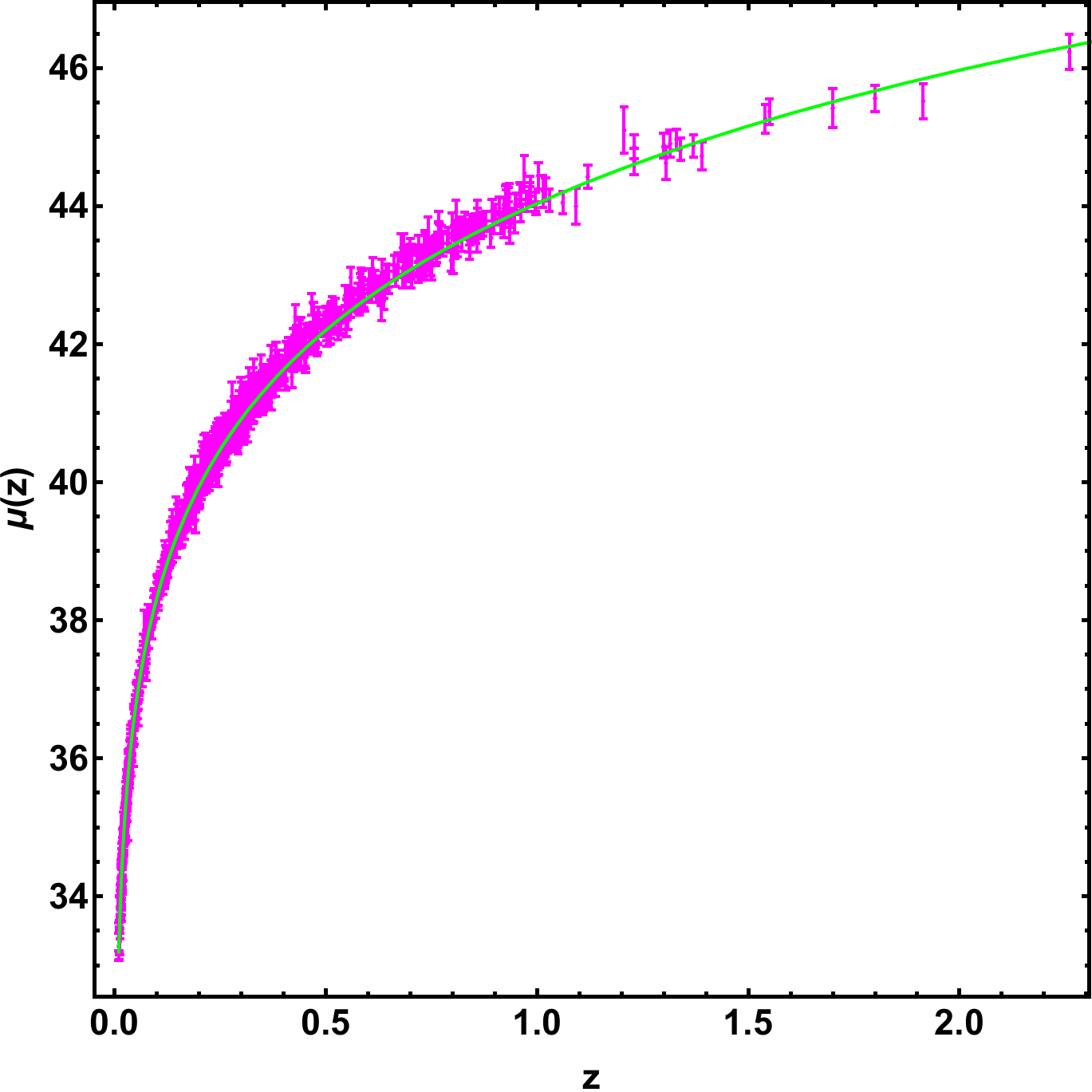}
\caption{Comparison between the observational Pantheon data and the theoretical distance modulus (green line) for the model $\displaystyle k=\frac{3}{2}$,
corresponding to the best fit values for the parameters.}
\label{moduluspantheonk32}
\end{figure}

The SNeIa sample consists of $1048$ objects in the range $0.01 < z < 2.26$. This sample is a combination of $365$ spectroscopically confirmed SNeIa discovered by the Pan-STARRS1 PS1 Medium Deep Survey together to the subset of $279$ PS$1$ SNeIa in the range ($0.03 < z < 0.68$) with distance estimates from SDSS, SNLS, various low redshift and HST samples. See \cite{scolnic2018} for details on photometry, astrometry, calibration, and systematic uncertainties.
The SNeIa observations provide the apparent magnitude $m(z)$ at the peak  
luminosity after several corrections.
The resulting apparent magnitude $m(z)$ can be easily related to
the Hubble free luminosity distance through the relation:
\begin{equation}
m_{th}(z)={\bar M}+ 5 \log_{10} (D_L (z))\,. \label{mdl}
\end{equation}
Here ${\bar M}$ is the zero point offset and depends on the absolute magnitude
$M$ and on the present value of the Hubble parameter. The cosmological model parameters can be  determined by minimizing
the quantity
\begin{equation}
\chi^2_{SNIa} ( \{\theta_{p}\})= \sum_{i=1}^N \frac{(\mu_{obs}(z_i) -
\mu_{th}(z_i,\{\theta_{i}\}))^2}{\sigma_{\mu \; i}^2 }\,.\label{chi2}
\end{equation}
The theoretical distance modulus is
therefore defined as
\begin{equation}
\mu_{th}(z_i,\{\theta_{p}\}) =5 \log_{10} (D_L(z_i,\{\theta_{p}\})) +\nu_0\,,
\end{equation}
where $D_L$ is the luminosity distance:
\begin{equation}
D_L =\frac{c}{100 h} (1+z)\int^{z}_{0}\frac{1}{H(\zeta, \theta)}d\zeta\,. 
\end{equation}
The parameter $\nu_0$  encodes the Hubble constant. The absolute magnitude $M$ and$\{\theta_{p}\}$ are model parameters. Actually, it is well known that using only SNeIa, one cannot constrain the Hubble constant, without including measurements of its local value from the SHOES project \cite{shoes,shoes2}, since this is degenerate with $M$. In order to set
the starting points for our chains, we first perform a preliminary and standard fitting procedure to maximize the
likelihood function ${\cal{L}}({\bf p})$:
\begin{eqnarray}
{\mathcal{L}}({\bf p}) & \propto & \frac {\exp{(-\chi^2_{SNIa}/2)}}{(2 \pi)^{\frac{{\cal{N}}_{SNIa}}{2}} |{\bf C}_{SNIa/GRB}|^{1/2}}\,.
\label{defchiall}
\end{eqnarray}
To build up their own regions of confidence,  we use the Bayesian approach based on  the Markov
Chain Monte Carlo (MCMC) method.  According to this procedure, we
 run three parallel chains and use the Gelman - Rubin diagnostic approach to test the convergence. We discard the first $30\%$ of the point iterations at the beginning of any MCMC run, and thin the chains that have to be  run many times. We finally extract the constrains on the parameters by co-adding the
thinned chains. It turns out that $H_0=0.98\pm 0.07$\,,$u_0=0.23\pm 0.09$\,, and $\Sigma=1.1\pm 0.2$. In Fig. (\ref{moduluspantheonk32}),  we plot the best fit curve with superimposed  data set.


\section{Discussion and Conclusions}
\label{conclusion}
In this paper, we investigated the possibility that coupled quintessence dynamics could be derived from the Noether Symmetry Approach. The method allows to select both the self-interacting potential of the scalar field and the analytical  form of the interaction. We recover the exponential  function widely used in literature, just confirming that  the existence of the Noether symmetries offers a physical criterion to fix  potentials, couplings and, in general, the form of models. Such an approach revealed extremely useful also for other classes of models \cite{RNCIB,leandros,felixGB}. Moreover, by this approach, we are able to solve exactly the Friedmann equations, at least in the case of dark energy and matter dominated Universe. Specifically, we obtain a one-parameter family of exact solutions. The main cosmological observables can be directly derived starting from the general solutions. 
In order to match the results with observations, we consider a special value of the parameter $k$, and compared the theoretical solution  with  observational dataset using Pantheon data on SNeIa. It turns out, that the model is quite well compatible with this SNeIa data set.
In forthcoming paper,  we will discuss solutions coming from Noether symmetries with different datasets in order to achieve a reliable cosmic history at different redshifts.   
\section*{Acknowledgements}
We  acknowledge INFN Sez. di Napoli ({\it Iniziative Specifiche} MOONLIGHT$2$, QGSKY and TEONGRAV). 
This article is based upon work from COST Action CA$15117$ Cosmology and Astrophysics 
Network for Theoretical Advances and Training Actions (CANTATA), 
supported by COST (European Cooperation in Science and Technology). 
We acknowledge the anonymous referee for her/his suggestions that allowed  to improve the manuscript.

\begin{onecolumn}
\appendix
\section{Cosmological solutions for  the cases $k=-3,\,3/2,\,3$}
\label{Appendix:A}

In this Appendix we show the expressions for the scale factor and the scalar field corresponding to the cases we considered. 
\begin{itemize}
\item  $k=-3$. It is:
\begin{eqnarray}
 &&a(t)^3=\frac{t \left(3 \Sigma  t+8 u_0\right)}{2 \left(3 \Sigma +4 u_0\right) \left(3 \Sigma +8
   u_0\right){}^2}\times\nonumber\\
   &&\left[9 \Sigma ^2 \left(\left(3 H_0-2\right) t^2-3 H_0+4\right)+24 \Sigma  u_0 \left(3 H_0
   \left(t^2+t-2\right)-t (t+2)+6\right)\right.\nonumber\\
   &&\left.+64 u_0^2 \left(3 H_0 (t-1)-t+2\right)\right]\,,\label{eqaminus3}
\end{eqnarray}
\begin{eqnarray}
&&\exp\left[\sqrt{\frac{3}{2}}\phi(t)\right]=\frac{1}{\left(3 \Sigma +4 u_0\right) \left(3 \Sigma +8 u_0\right){}^2 \left(3 \Sigma 
   t+4 u_0\right){}^2}\times\nonumber\\&&\left[8 t \left(3 \Sigma  t+8 u_0\right) \left(3 H_0 (t-1) \left(3 \Sigma +8 u_0\right)
   \left(3 \Sigma  (t+1)+8 u_0\right)\right.\right.\nonumber\\
   &&\left.\left.-2 \left(3 \Sigma +4 u_0\right) \left(3 \Sigma 
   \left(t^2-2\right)+8 (t-2) u_0\right)\right)\right]\,.\label{eqphiminus3}
\end{eqnarray}


\item $k=3/2$. It is 
\begin{eqnarray}
&& a(t)^3= \frac{1}{\left(3 \Sigma +4 u_0\right){}^2 \left(3 \Sigma ^2+16 \Sigma \left(u_0+1\right)+16 u_0 \left(u_0+2\right)\right)}\times\nonumber\\
&& \left\{ t \left[27 \Sigma ^4 t^2 \left(3 H_0 (t-1)-3 t+4\right)+36 \Sigma ^3 t \left(3
   u_0+2\right) \left(\left(3 H_0-2\right) t^2-3 H_0+4\right)\right.\right.\nonumber\\ 
   &&\left.\left.+64 \Sigma  u_0^2 \left(5
   u_0+8\right) \left(3 H_0 \left(t^2-1\right)-t^2+3\right)+96 \Sigma ^2 u_0 \left(3
   H_0 (t-1) \left(t \left((t+4) u_0+t+5\right)+u_0+1\right)\right.\right.\right. \nonumber\\
   &&\left.\left.\left. +4 \left(3 t+u_0+1\right)-t
   \left(t \left((t+6) u_0+t+8\right)-9 u_0\right)\right)+256 u_0^3 \left(u_0+2\right)
   \left(3 H_0 (t-1)-t+2\right)\right]\right\},
   \label{eqa3over2}
   \end{eqnarray}
  \begin{eqnarray}
&&\exp\left[\sqrt{\frac{3}{2}}\phi(t)\right]=\frac{1}{\left(3 \Sigma +4 u_0\right){}^2 \left(\Sigma  (3 \Sigma +16)+16 u_0 \left(\Sigma
   +u_0+2\right)\right) \left(3 \Sigma  t+4 u_0\right)}\times\nonumber\\
 && 16 t \left[3 H_0 (t-1) \left(3 \Sigma +4 u_0\right) \left(\Sigma  ((3 \Sigma +8) t+8)+8
   u_0 \left(\Sigma +\Sigma  t+2 u_0+4\right)\right)\right.\nonumber\\
   &&\left. +3 \Sigma ^2 \left(-(9 \Sigma +16)
   t^2+12 \Sigma  t+32\right)+4 u_0 \left(\Sigma  (24 (\Sigma +4)-t (-9 \Sigma +2 (9
   \Sigma +4) t+48))\right.\right.\nonumber\\&&\left.\left.+4 u_0 \left(15 \Sigma -2 t (\Sigma  (t+3)+4)-4 (t-2)
   u_0+16\right)\right)\right]\,.
\label{phi3over2}
\end{eqnarray}

\item $k=3$. It is 
\begin{eqnarray}
&& a(t)^3=\frac{t \left(3 \Sigma  t+4 u_0\right)}{2 \left(\Sigma +2 u_0\right) \left(3 \Sigma +4
   u_0\right){}^2}\times\nonumber\\
   && 3 \Sigma ^2 \left[\left(\left(3 H_0-2\right) t^2-3 H_0+4\right)+4 \Sigma  u_0 \left(3 H_0
   (t-1) (t+4)\right.\right.\nonumber\\
   &&\left.-t (t+6)+12\right)+16 u_0^2 \left(3 H_0 (t-1)-t+2\right]\,.
   \label{eqak3}\\ 
&&\exp\left[\sqrt{\frac{3}{2}}\phi(t)\right]=\frac{8 t}{\left(\Sigma +2 u_0\right) \left(3 \Sigma +4 u_0\right){}^2
   \left(3 \Sigma  t+4 u_0\right)}  \times\nonumber\\
   && \left[3 \Sigma ^2 \left(\left(3 H_0-2\right) t^2-3 H_0+4\right)+4 \Sigma  u_0
   \left(3 H_0 (t-1) (t+4)-t (t+6)+12\right)\right.\nonumber\\
   &&\left. +16 u_0^2 \left(3 H_0
   (t-1)-t+2\right)\right]\,.
\label{eqphik3}
\end{eqnarray}
\end{itemize}

\end{onecolumn}

\end{document}